\documentclass{svjour3}
\usepackage{graphicx}
\usepackage{amsfonts}
\usepackage{amsmath}

\begin{document}

\title{Spin-structures of N-boson systems with nonzero spins}
\subtitle{- an analytically solvable model with pairing force}

\author{C.G. Bao}

\institute{ State Key Laboratory of Optoelectronic Materials and
Technologies,
 Sun Yat-sen University,
 Guangzhou, 510275, P.R. China \\
 \email{stsbcg@mail.sysu.edu.cn \\
 Correspondence: C.G. Bao, Department of Physics, Sun Yat-sen University, Guangzhou, 510275, P.R. China}
}

\date{Received: date / Accepted: date}

\maketitle

\begin{abstract}
A model is proposed to study the possible pairing structures of N-boson
systems with nonzero spin. Analytical solutions have been obtained. The
emphasis is placed on the spin-structures of ground states with attractive
or repulsive pairing force, and with or without the action of a magnetic
field. A quantity (an analogue of the two-body density function) is defined
to study the spin-correlation between two bosons in N-body systems. The
excitation of the system has also been studied.
\end{abstract}

\section{Introduction}

Historically, the pairing structures played a very important role in fermion
systems. The most famous example is the Cooper-pair in superconductivity.
\cite{BJ1957} One more example is the pairing of the valence nucleons in
nuclei as described by the "Interacting Boson Model" \cite{AA1976}. However,
there is no obvious evidence to support the existence of paring structures
in boson systems. Nonetheless, we can not rule out the possibility. For an
example, each $^{52}$Cr atom has a spin $J=3$. The S-wave scattering lengths
$a_{s}$ of these atoms depend on the spin-channels $s$, namely, the total
spin of the two interacting atoms. There are four channels, $s=0,2,4,$ and $%
6 $. Accordingly, for low energy scattering, the interaction between the $i$
and $j$\ atoms can be written as
\begin{equation}
v_{ij}=\delta (\mathbf{r}_{i}-\mathbf{r}_{j})\sum_{s}g_{s}\mathfrak{P}%
_{ij}^{s}  \label{eq:01}
\end{equation}%
where the strength $g_{s}=4\pi \hbar ^{2}a_{s}/M$, $M$ is the mass of atom. $%
\mathfrak{P}_{ij}^{s}$\ is the projection operator of the $s-$channel. Let $%
\chi (i)$\ be the spin-state of $i$, $(\chi (i)\chi (j))_{sm}\equiv
(ij)_{sm} $ be the normalized total spin-state of $i$ and $j$ , where the
spins are coupled to $s$ and $m$\ (the Z-component of $s$) via the
Clebsch-Gordan coefficients. Then, $\mathfrak{P}_{ij}^{s}\equiv
\sum_{m}|(ij)_{sm}\rangle \langle (ij)_{sm}|$. For $^{52}$Cr, it has been
determined experimentally that $a_{2}=-7a_{B},$ $a_{4}=58a_{B},$ and $%
a_{6}=112a_{B}$, while $a_{0}$ is undetermined.\cite{SJ2005} Since $g_{4}$
and $g_{6}$ \ are strongly repulsive, the atoms in low-lying states would do
their best to avoid forming $s=4$\ and 6 pairs. Although $g_{2}$\ is
attractive, it is weak. Therefore, if $g_{0}$\ is also attractive and
strong, the low-lying states might be dominated by the force of the $s=0$
channel (the pairing force) to form $s=0$\ pairs. Therefore, in view of
having so many different kinds of boson systems (the constituent may be an
atom, a pair of fermions, an exciton, etc.), the assumption that some
boson-systems in some cases might be dominated by pairing force is assumable.

In what follows, we shall propose a model to study the effect of the pairing
force on boson systems. In this model, each particle has a nonzero integral
spin $J$, and they are interacting via pairing interaction. In general, if
one can obtain analytical solutions of a Hamiltonian, then these solutions
are valuable because the underlying physics can be thereby better
understood. Besides, as an exact solution, it can be used to check the
validity of various approximate methods. It turns out that the above model
can be solved analytically as shown below.

Since the experimental realization of the condensation of atoms with nonzero
spins,\cite{ST} the study of the spinor condensates has become a hot topic
due to academic interest and \ their potential in application.
Traditionally, the mean-field theory is used to explain related phenomena.
Up to now this theory is successful. Nonetheless, if the details of
spin-spin correlation are taken into account, this theory is not sufficient
due to the basic approximation inherent in the theory. On the other hand,
the theoretical method used in this paper is beyond the mean-field theory.
In what follows, approximation is made only in the spatial space, while the
spin-degrees of freedom of the $N-$body system have been exactly treated.
Thus, the following approach would be a useful complement to the mean-field
theory. The spin-spin correlation might by thereby better understood.

\section{Hamiltonian, spectrum, and the eigenstates}

We first consider a N-boson system, each boson has spin $J=3$. The cases
with $J\neq 3$\ will be discussed later. The bosons are confined by a
potential. The interaction acting among the bosons contains a
spin-independent term and a spin-dependent term. The latter is simply the
pairing force, namely,
\begin{equation}
v_{ij}=\delta (\mathbf{r}_{i}-\mathbf{r}_{j})g_{0}\mathfrak{P}_{ij}^{0}\ .
\label{eq:02}
\end{equation}

We are only interested in the case of the low-temperature limit. At this
limit all particles are condensed into the same spatial state $f(\mathbf{r)}$%
.\cite{REF4} In other words, the spatial state is fixed. Therefore, the
spatial degrees of freedom can be integrated. After the integration, the
Hamiltonian of the model reads
\begin{equation}
H_{mod}=\sum_{i<j}G_{0}\mathfrak{P}_{ij}^{0}  \label{eq:03}
\end{equation}%
where $G_{0}\equiv g_{0}\int d\mathbf{r}|f(\mathbf{r})|^{4}.$ For this
Hamiltonian, only the spin-degrees of freedom are considered. Obviously, $f(%
\mathbf{r})$ depends on the potential of confinement and on the interaction.
However, since $G_{0}$ will be considered as a parameter, the details of the
potential and the spin-independent interaction is irrelevant.\ In what
follows, $g_{0}$ (and therefore $G_{0}$) is first considered to be negative.

For the diagonalization of $H_{mod}$, we introduce the set of symmetrized
and normalized Fock-states $|k\rangle \equiv
|N_{3}^{k},N_{2}^{k},N_{1}^{k},N_{0}^{k},N_{-1}^{k},N_{-2}^{k},N_{-3}^{k}%
\rangle $ as basis functions, where $N_{\mu }^{k}$\ is the number of atoms
with spin component $\mu $, and $k$\ denotes the set $\{N_{\mu }^{k}\}$
which together determine the basis function. Obviously, $\sum_{\mu }N_{\mu
}^{k}=N,$ the total number of particles. Since only the spin-degrees of
freedom are involved in $H_{mod},$ the set of Fock-states is complete. The
related matrix elements read
\begin{eqnarray}
\langle k^{\prime }|H_{mod}|k\rangle &=&\frac{G_{0}}{14}\sum_{\mu ,\nu
}(-1)^{\mu +\nu }\sqrt{N_{\mu }^{k}N_{\nu }^{k^{\prime }}}[\ \delta _{\nu
,0}\delta _{\mu ,0}\delta _{k^{\prime }k}(N_{0}^{k}-1)  \notag \\
&&\ +\overline{\delta }_{\nu ,0}\delta _{\mu ,0}\sqrt{N_{-\nu }^{k^{\prime
}}(N_{0}^{k}-1)}\delta _{N_{0}^{k}-2}^{N_{\nu }^{k^{\prime }}-1,N_{-\nu
}^{k^{\prime }}-1}  \notag \\
&&\ +\delta _{\nu ,0}\overline{\delta }_{\mu ,0}\sqrt{(N_{0}^{k^{\prime
}}-1)N_{-\mu }^{k}}\delta _{N_{\mu }^{k}-1,N_{-\mu
}^{k}-1}^{N_{0}^{k^{\prime }}-2}  \notag \\
&&\ +\overline{\delta }_{\nu ,0}\overline{\delta }_{\mu ,0}\sqrt{N_{-\nu
}^{k^{\prime }}N_{-\mu }^{k}}\delta _{N_{\mu }^{k}-1,N_{-\mu
}^{k}-1}^{N_{\nu }^{k^{\prime }}-1,N_{-\nu }^{k^{\prime }}-1}\ ]
\label{eq:04}
\end{eqnarray}%
where $\overline{\delta }_{\mu ,\nu }\equiv (1-\delta _{\mu ,\nu })$. For
the label $\delta _{N_{0}^{k}-2}^{N_{\nu }^{k^{\prime }}-1,N_{-\nu
}^{k^{\prime }}-1}$, the superscript implies a revised set of the set $%
\{N_{3}^{k^{\prime }}\cdots N_{-3}^{k^{\prime }}\}$ by reducing both $N_{\nu
}^{k^{\prime }}$ and $N_{-\nu }^{k^{\prime }}$ by 1, the subscript implies a
revised set of $\{N_{3}^{k}\cdots N_{-3}^{k}\}$ by reducing $N_{0}^{k}$ by
2. When the two revised sets are one-to-one identical, the label is 1,
otherwise it is zero. And so on. When $N$\ is given, the dimension of the
matrix is finite. Due to the special structure of the matrix, it can be
analytically diagonalized. The results are the follows.

For convenience, $N$ is assumed to be even unless specified. The ground
state $\Psi _{g}$ has an analytical form as
\begin{eqnarray}
\Psi _{g} &=&\sum_{k}C_{k}^{g}|k\rangle ,\ \ \ \mbox{(not yet normalized)}
\label{eq:05} \\
C_{k}^{g} &=&\delta _{N_{3}^{k},N_{-3}^{k}}\delta
_{N_{2}^{k},N_{-2}^{k}}\delta
_{N_{1}^{k},N_{-1}^{k}}(-1)^{N_{3}^{k}+N_{1}^{k}}\sqrt{N_{0}^{k}!}%
/((N_{0}^{k}/2)!\ 2^{N_{0}^{k}/2})  \label{eq:06}
\end{eqnarray}%
where $N_{0}^{k}$\ should be even. On the other hand, we define
\begin{equation}
\Psi _{0}^{pair}\equiv \boldsymbol{P}\{(12)_{0}(34)_{0}\cdots (N-1,N)_{0}\}
\label{eq:07}
\end{equation}%
where $\boldsymbol{P}$ implies a summation over all the $N!$ permutations of
particles, and the implication of $(ij)_{0}$ has been given in the previous
section.\ This state can be expanded in terms of the Fock-states as
\begin{equation}
\Psi _{0}^{pair}\equiv \frac{(-1)^{N/2}}{7^{N/4}}(N/2)!2^{N/2}\sqrt{N!}%
\sum_{k}C_{k}^{g}|k\rangle \ .  \label{eq:08}
\end{equation}%
Hence, $\Psi _{0}^{pair}$ is just the ground state that we have obtained
from the diagonalization. Thus the structure of the ground state is clear,
it is just a product of the $s=0$ pairs, i.e., a pairing structure.

We use $|G_{0}|$\ as the unit of energy in the follows, then the ground
state energy turns out to be $\ \ \ \ \ \ \ \ \ \ \ \ \ E_{g}=-\frac{N(N+5)}{%
14}\equiv \varepsilon _{N}$.

If $N$\ is odd, the ground state would be just the above pairing structure
together with an additional particle, and $E_{g}=\varepsilon _{N}+3/7$.

It is reminded that the interaction keeps the total spin $S$\ and its
Z-component $M$\ to be conserved. Obviously, due to the pairing structure,
the ground state has $S=0$. The first excited level is found to contain the
eigen-states having $S=$ 2, 4, and 6. They are exactly degenerate. Their
analytical forms read
\begin{eqnarray}
\Psi _{y,SM} &=&\sum_{k}C_{k}^{y,SM}|k\rangle ,\ \ \
\mbox{(not yet
normalized)}  \label{eq:09} \\
C_{k}^{y,SM} &=&\sum_{\nu }C_{3,\nu ,3,M-\nu }^{SM}F_{k\nu }  \label{eq:10}
\end{eqnarray}%
where the script $y$ implies that the state is a yrast state, and the
Clebsch-Gordan coefficients have been introduced, and
\begin{eqnarray}
F_{k\nu } &=&\delta _{L_{3},L_{-3}}\delta _{L_{2},L_{-2}}\delta
_{L_{1},L_{-1}}(-1)^{L_{3}+L_{1}}\sqrt{L_{0}!}/((L_{0}/2)!2^{L_{0}/2})
\notag \\
&&\times \lbrack \ \delta _{\nu ,M-\nu }\sqrt{(L_{\nu }+2)(L_{\nu }+1)}+%
\overline{\delta }_{\nu ,M-\nu }\sqrt{(L_{\nu }+1)(L_{M-\nu }+1)}\ ]
\label{eq:11}
\end{eqnarray}%
where $L_{\mu }$ and $N_{\mu }^{k}$ are related as $L_{\mu }=N_{\mu
}^{k}-\delta _{\mu ,\nu }-\delta _{\mu ,M-\nu }\ $( $\mu $\ is from -3 to 3)
and $L_{0}$ should be even.

Similarly, we define an excited pairing structure by changing a $s=0$\ pair
to a $s\neq 0$ pair. We found
\begin{equation}
\boldsymbol{P}\{(12)_{SM}(34)_{0}\cdots (N-1,N)_{0}\}\equiv cons\times
\sum_{k}C_{k}^{y,SM}|k\rangle \ .  \label{eq:12}
\end{equation}%
Thus, it is clear that the $\Psi _{y,SM}$ states have just the excited
pairing structures. They all have the same eigenenergy $E_{g}+1$
disregarding $S$ and $N,$ and the gap is just 1 (in $|G_{0}|$, \cite{REF5} ).

In general, the spectrum of this $N-$body system contains a finite number of
levels, each is highly degenerate (except the lowest one).\ \ We introduce a
quantum number $I$\ to denote the levels, therefore $I$\ describes the
degree of excitation. Each eigenenergy is a sum of two terms, it reads
\begin{equation}
E_{I}=\varepsilon _{N}+\varepsilon _{I}\ .  \label{eq:13}
\end{equation}%
Where $\varepsilon _{N}=-\frac{N(N+5)}{14}$ depends roughly on the number of
pairs ($\sim N^{2}$) and is the same for all the eigen-states, while $%
\varepsilon _{I}=\frac{I(I+5)}{14}$ depends on $I$, $I=0,2,4,\cdots N$ (if $N
$ is even, in this case the first excited level has $I=2$\ and therefore $%
\varepsilon _{I}=1$), or $I=1,3,\cdots N$ (if $N$ is odd). So, when $N$ is
even (odd), totally there are $N/2+1$\ (($N$+1)/2) energy levels. When $I$
is larger, the level will be not only higher, but also have a larger
degeneracy as shown later. When we compare the spectra of two systems
distinct in $N,$\ their $\varepsilon _{I}$ are one-to-one identical (except
the range of $I$). Thus, although the spectrum of the one with more
particles would contain more levels, the lower parts of their spectra are
exactly the same except the shift caused by $\varepsilon _{N}$. This is a
noticeable point. The spacing of adjacent levels is just $%
E_{I+2}-E_{I}=(2I+7)/7$. Thus, when $I$\ goes up, the spacing becomes larger
and larger. Note that, since the set of Fock-states is complete, the above
spectrum is exact for the model Hamiltonian.

For all the states of the I-level, $I$ atoms would be excited from the $s=0$
pairs. Therefore, $I$ has a similar implication as the seniority which has
been introduced for describing the excitation of pairing structures of
fermion systems long ago.\cite{APW1958,KAK1961} Furthermore, the maximum $S$%
\ of the states belonging to an I-level is obvious $3I$ contributed by the $%
I $ excited atoms.

In general, the eigen-states can be classified according to $S$, $M$, and $I$
. It is possible that there are more than one states having the same set $%
(ISM)$ (e.g., for the $I=4$ level, there are two independent $S=4$ \
states). In this case an extra quantum number is necessary. Therefore, an
eigenstate can be labeled as $\Psi _{I,S,M,i}$, where $i$\ denotes other
good quantum numbers, if necessary. For examples, the above three yrast
states $\Psi _{y,SM}$\ have $I=2$, therefore they can also be denoted as $%
\Psi _{2,SM}$.

The degeneracy of an I-level is stated as follows. Let $\{j_{\mu }\}$\ be a
set of seven non-negative integers ($\mu $ is from -3 to 3). There would be
a finite number of sets satisfy both $\sum_{\mu }j_{\mu }=I$ and $\sum_{\mu
}\mu j_{\mu }=0$. The number of these sets is denoted by $\Omega _{I}$.
Then, one can prove that the degeneracy of the I-th level $d_{I}=\Omega
_{I}-\Omega _{I-2}$ ($d_{0}=d_{1}=1$). This number increases very fast with $%
I$, e.g., $d_{0}=1$, $d_{2}=3$, and $d_{4}=14$. Note that the degeneracy
caused by $M$ has not yet been taken into account.

For higher excitation, a stste from the $I=4$ level is given as an example
as
\begin{equation}
\Psi _{400}=cons\times (3\boldsymbol{P}\{[(12)_{2}(34)_{2}]_{0}(56)_{0}%
\cdots \}-\sqrt{5}\boldsymbol{P}\{[(12)_{4}(34)_{4}]_{0}(56)_{0}\cdots \})
\label{eq:14}
\end{equation}%
where four particles are excited from the $s=0$ pairs. The structures of
higher states will become more and more complicated because more particles
are excited. We are not going to the details of them.

When $G_{0}$ is positive, the above analytical solutions hold exactly.
However, the spectrum would be reversed. The ground states would be highly
degenerate, all of them belong to the $I=N$ level, wherein all $s=0$ pairs
are rejected. In particular, there is a great energy gap $%
E_{gap}=E_{N-2}-E_{N}=(2N+3)/7$. A larger $N$\ leads to a larger gap. Thus,
when $N$\ is large, the spin-degrees of freedom of the group of ground
states are difficult to be excited. This is also a noticeable point.

When $J\neq 3$, one can also use the Fock-states as basis functions. After
the diagonalization, the eigenenergies are also a sum of two terms, namely, $%
E_{I}=\varepsilon _{N}+\varepsilon _{I}$. However, $\varepsilon _{N}=-\frac{
N(N+2J-1)}{2(2J+1)}$ , while $\varepsilon _{I}=\frac{I(I+2J-1)}{2(2J+1)}$.
The eigenstates have the same structures as before, and therefore can also
be classified as $\Psi _{ISMi}$. In particular, the $I=0$ level has all
particles in $s=0$\ pairs, while the $I=2$\ level has two particles excited
from the $s=0$ pairs, and so on.

\section{Ground states under a magnetic field}

We go back again to the cases with $J=3$\ and an even $N$ and a negative $%
G_{0}$. When a magnetic field $B$ lying along the Z-axis is applied, the
model Hamiltonian reads (in $|G_{0}|$)
\begin{equation}
H_{mod}^{\prime }=-\sum_{i<j}\mathfrak{P}_{ij}^{0}-\gamma \hat{S_{Z}}
\label{eq:15}
\end{equation}%
where $\hat{S_{Z}}$ is the operator of the total spin along the Z-axis, $%
\gamma =g\mu _{B}B/|G_{0}|$, $\mu _{B}$ is the Bohr magneton (incidentally,
if the atom is $^{52}$Cr with $J=3$, $g=2$). The eigen-states of $%
H_{mod}^{\prime }$ are exactly the same as $H_{mod}$, however the
eigen-energies would contain an additional term due to the Stern-Gerlach
splitting as
\begin{equation}
E_{ISM}^{\prime }=\varepsilon _{N}+\varepsilon _{I}-\gamma M\ .
\label{eq:16}
\end{equation}

We are interested in how the structure of the ground state would vary with $B
$. For the group of states belonging to the $I-$level, the largest $M$
possessed by these states is$\ 3I$. Thus the state with $S=M=3I$\ of each $I-
$level is a candidate of the ground state with the energy $%
E_{I,3I,3I}^{\prime }$. \ Among the candidates, when $B$\ (or $\gamma $) is
fixed, if $I=I_{o}$ leads to the minimum of the group $E_{I,3I,3I}^{\prime }$
,\ then the ground state would have $I=I_{o}$. Obviously, $I_{o}$ depends on
$B$. It was found that, when $B$\ is small so that $\gamma <7/42$, the
ground state has $I_{o}=0$. Starting from 7/42, let the range of $\gamma $\
be divided into segments, each has a length 2/21. Then, when $\gamma $\
increases, $I_{o}$ will increase step by step from a segment to the next
segment. Each step $I_{o}$ will jump by 2.\ The general relation of $I_{o}$\
and $\gamma $ is:
\begin{equation}
(2I_{o}+3)/42\leq \gamma <(2I_{o}+7)/42\ .  \label{eq:17}
\end{equation}

Since $I_{o}$ can not exceed $N$, the last segment is $(2N+3)/42\leq \gamma
<\infty .$\ In this extensive segment, \thinspace $I_{o}=N$, namely, the
increase of the magnetic field will eventually excite all particles from the
$s=0$ pairs.

In each segment the associated ground state energy $E_{gr}(\gamma
)=E_{I_{o},3I_{o},3I_{o}}^{\prime }$, and $\frac{d}{d\gamma }E_{gr}(\gamma
)=-3I_{o}$ within the segment. When $\gamma $\ increases, since $I_{o}$ will
increase step by step, the slope becomes more and more negative.

When $\gamma <(2N+3)/42,$ we have approximately $\gamma \approx (2I_{o}+5)/42
$. Thus we have $E_{gr}(\gamma )\approx -\varepsilon _{N}-(42\gamma
-5)^{2}/56.$ It implies that, when $\gamma $ increases, the decline of $%
E_{gr}(\gamma )$ against $\gamma $\ is nearly parabolic. Of course, when $%
\gamma >(2N+3)/42,$ the decline is only contributed by the last term of
eq.(16), i.e., $-3\gamma N$, and therefore is linear.

The analytical form of the wave function of the ground state can also be
obtained. It reads in terms of the Fock-states as (not yet normalized)
\begin{eqnarray}
\Psi _{gr}(\gamma ) &\equiv &\Psi _{I_{o},3I_{o},3I_{o}}=\sum_{k}\delta
_{N_{3}^{k}-I_{o},N_{-3}^{k}}\delta _{N_{2}^{k},N_{-2}^{k}}\delta
_{N_{1}^{k},N_{-1}^{k}}(-1)^{N_{3}^{k}+N_{1}^{k}}\sqrt{%
N_{0}^{k}!N_{3}^{k}!/N_{-3}^{k}!}  \notag \\
&&\ \ \ \ \ \ \ \ \ \ \ \ \ \ \ \ \ \ \ \ \ /((N_{0}^{k}/2)!\
2^{N_{0}^{k}/2})\ |k\rangle   \notag \\
&=&cons\times \boldsymbol{P}\{[\Pi _{i=1}^{I_{o}}\chi
_{3}(i)](I_{o}+1,I_{o}+2)_{0}\cdots (N-1,N)_{0}\}\ .  \label{eq:18}
\end{eqnarray}

Thus, when a magnetic field is applied, the ground state would have a
portion of particles excited and become fully polarized, while the others
remain in the $s=0$ pairs. Thus the implication of $I_{o}$\ is clear, it is
just the number of polarized particles.

From $\Psi _{gr}(\gamma )$ it is straight forward to calculate the
probability of a particle in $\mu -$component $P_{\mu }\equiv \langle \Psi |%
\hat{N_{\mu }}|\Psi \rangle /N$, where $\hat{N_{\mu }}$ is the operator of
the number of particles in $\mu ,$ and $\sum_{\mu }P_{\mu }=1$. Obviously,
if $P_{3}=1$\ (or $P_{-3}=1$) while the other $P_{\mu }=0$ , the system is
fully polarized. Therefore, from $P_{\mu }$ one can understand the situation
of magnetization of a state. For $\Psi _{gr}(\gamma )$, from (18) we have
\begin{eqnarray}
&&P_{3}=\frac{I_{o}(I_{o}+N+6)+N}{N(7+2I_{o})}  \notag \\
&&P_{-3}=\frac{(I_{o}+1)(N-I_{o})}{N(7+2I_{o})}  \notag \\
&&P_{\mu }=\frac{N-I_{o}}{N(7+2I_{o})}.\ \ \ \mbox{(if $\mu \neq \pm3$)}
\label{eq:19}
\end{eqnarray}

\begin{figure}[tbp]
\resizebox{0.95\columnwidth}{!}{\includegraphics{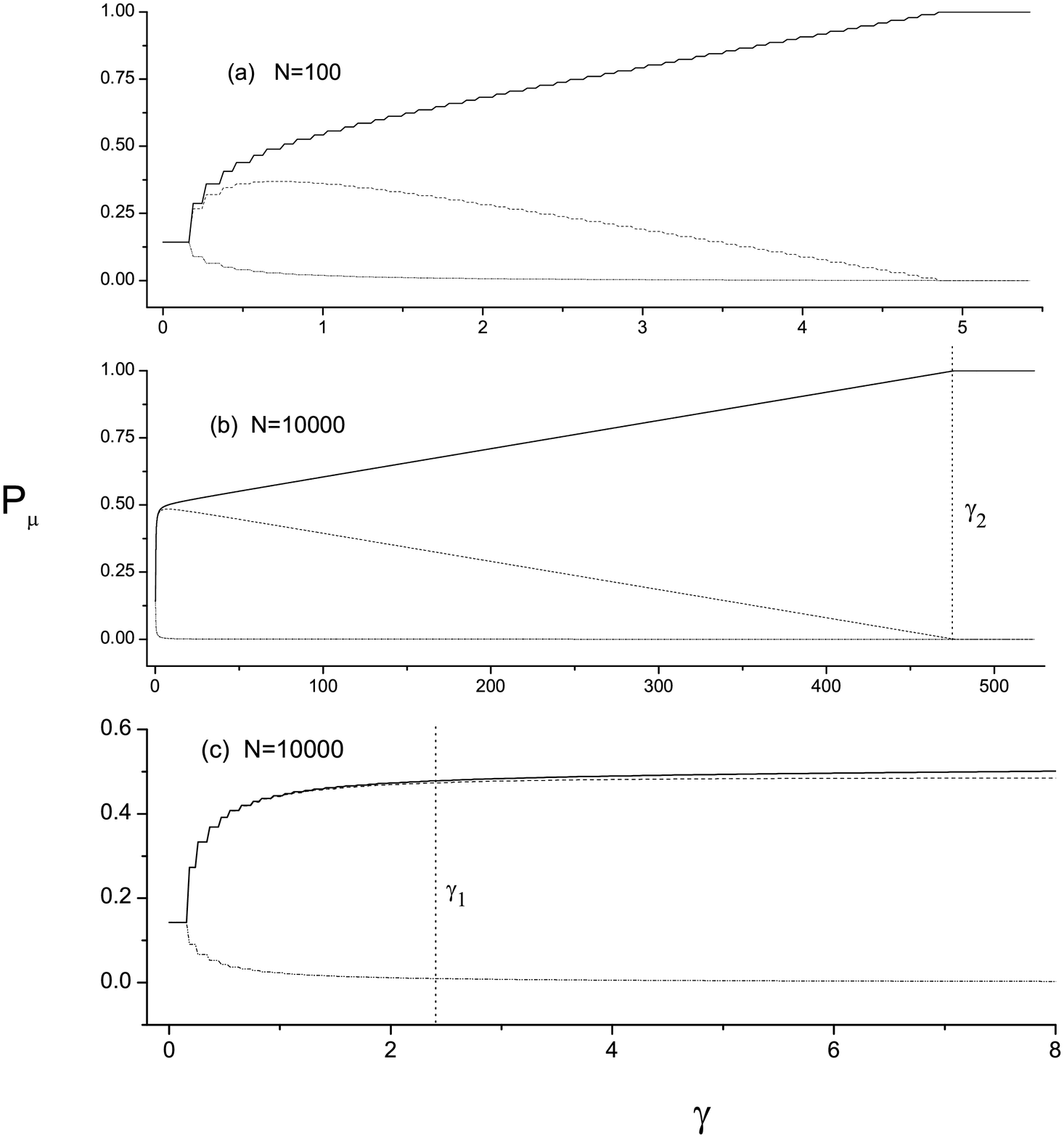}}
\caption{$P_{\protect\mu }$ (the probability of an atom in
$\protect\mu $
component) of the ground states of $H_{mod}^{\prime }$\ against $\protect%
\gamma $ with $N=100$ (a) and 10000 (b,c). $P_{3}$ is in solid line, $P_{-3}$
in dash line, and all other $P_{\protect\mu }$ are in dash-dot-dot line
(they overlap exactly). (c) is just the left end of (b).}
\label{fig:1}
\end{figure}

Examples of $P_{\mu }$\ against $\gamma $\ are given in Fig.1. When $\gamma
=0$, all $P_{\mu }=1/7.$ In this case all particles are in $s=0$ pairs.
Accordingly, the total spin-state has $S=0$ and therefore is isotropic. When
$\gamma $ increases, the variation of $P_{\mu }$ can be roughly divided into
three stages. In the first stage both $P_{3}$ and $P_{-3}$ increases very
fast while the other $P_{\mu }$ decreases very fast as shown in the figure.
When $\gamma =93N/(21(2N+100))\equiv \gamma _{1},$ $\ P_{\mu }\ (\mu \neq
\pm 3)$ are found to be equal to $0.01$. We use the range 0 to $\gamma _{1}$
to define the first stage. The second stage is from $\gamma _{1}$ to $\gamma
_{2}=(2N+3)/42$ (say, when $N=10^{4}$, $\gamma _{1}=$2.2 and $\gamma _{2}=$
476.3 as shown in Fig.1c and 1b). In this stage $P_{3}$\ increases ($P_{-3}$
decreases) nearly linearly until equal to 1 (0), while the other $P_{\mu }$
remain $\approx 0$. The third stage ($\gamma \geq \gamma _{2}$) is
characterized by $P_{3}=1$, i.e., the system is fully polarized.

Incidentally, the structure of the ground states of the $^{52}$Cr
condensates under a magnetic field has already been studied by a number of
authors using the mean field theory.\cite{DRB2006, SL2006, MH2007}
Obviously, the spinors from the mean field theory can be compared with $%
P_{\mu }$. When $g_{0}$\ is very negative, two phases "polar" and
"ferromagnetic" have been found. The associated spinors are ($\cos \theta
,0,0,0,0,0,\sin \theta )$ and (1,0,0,0,0,0,0), respectively. Obviously, the
former (latter) corresponds to the above second (third) stage of $P_{\mu }$.
From the correspondence we know that the underlying physics of the "polar
phase" is just a kind of structure wherein a portion of particles are fully
polarized while all other particles are in $s=0$ pairs.

Using the concept of the fractional parentage coefficients one can extract
any pair of particles, say $i$ and $j$, from a spin-state $\Psi
=\sum_{k}C_{k}|k\rangle $\ as
\begin{eqnarray}
\Psi  &=&\sum_{\mu ,\nu }\chi _{\mu }(i)\chi _{\nu }(j)[\ \delta _{\mu \nu
}\sum_{k}C_{k}\sqrt{\frac{N_{\mu }^{k}(N_{\mu }^{k}-1)}{N(N-1)}}|\cdots
,N_{\mu }^{k}-2,\cdots \rangle   \notag \\
&&+\overline{\delta }_{\mu \nu }\sum_{k}C_{k}\sqrt{\frac{N_{\mu }^{k}N_{\nu
}^{k}}{N(N-1)}}|\cdots ,N_{\mu }^{k}-1,\cdots ,N_{\nu }^{k}-1,\cdots \rangle
\ ]  \label{eq:20}
\end{eqnarray}%
where the two Fock-states written explicitly at the right are for $(N-2)$%
-body systems. The former originates from $|k\rangle $ by changing $N_{\mu
}^{k}$ to $N_{\mu }^{k}-2,$ while the latter by changing both $N_{\mu }^{k}$
and $N_{\nu }^{k}$ \ to $N_{\mu }^{k}-1$ and $N_{\nu }^{k}-1,$ respectively.
From this expansion it is straight forward to calculate how the
magnetization $m$ of a selected pair would be. Let $\mu $ and $\nu $ denote
the spin-components of the two selected atoms $i$ and $j$, and $m=\mu +\nu $%
. Then, the probability of the pair in $m$\ reads
\begin{equation}
Q_{m}=\sum_{\mu ,\nu }\delta _{\mu +\nu ,m}[\ \delta _{\mu \nu
}\sum_{k}C_{k}^{2}\frac{N_{\mu }^{k}(N_{\mu }^{k}-1)}{N(N-1)}+\overline{%
\delta }_{\mu \nu }\sum_{k}C_{k}^{2}\frac{N_{\mu }^{k}N_{\nu }^{k}}{N(N-1)}\
]\ .  \label{eq:21}
\end{equation}

\begin{figure}[tbp]
\resizebox{0.95\columnwidth}{!}{\includegraphics{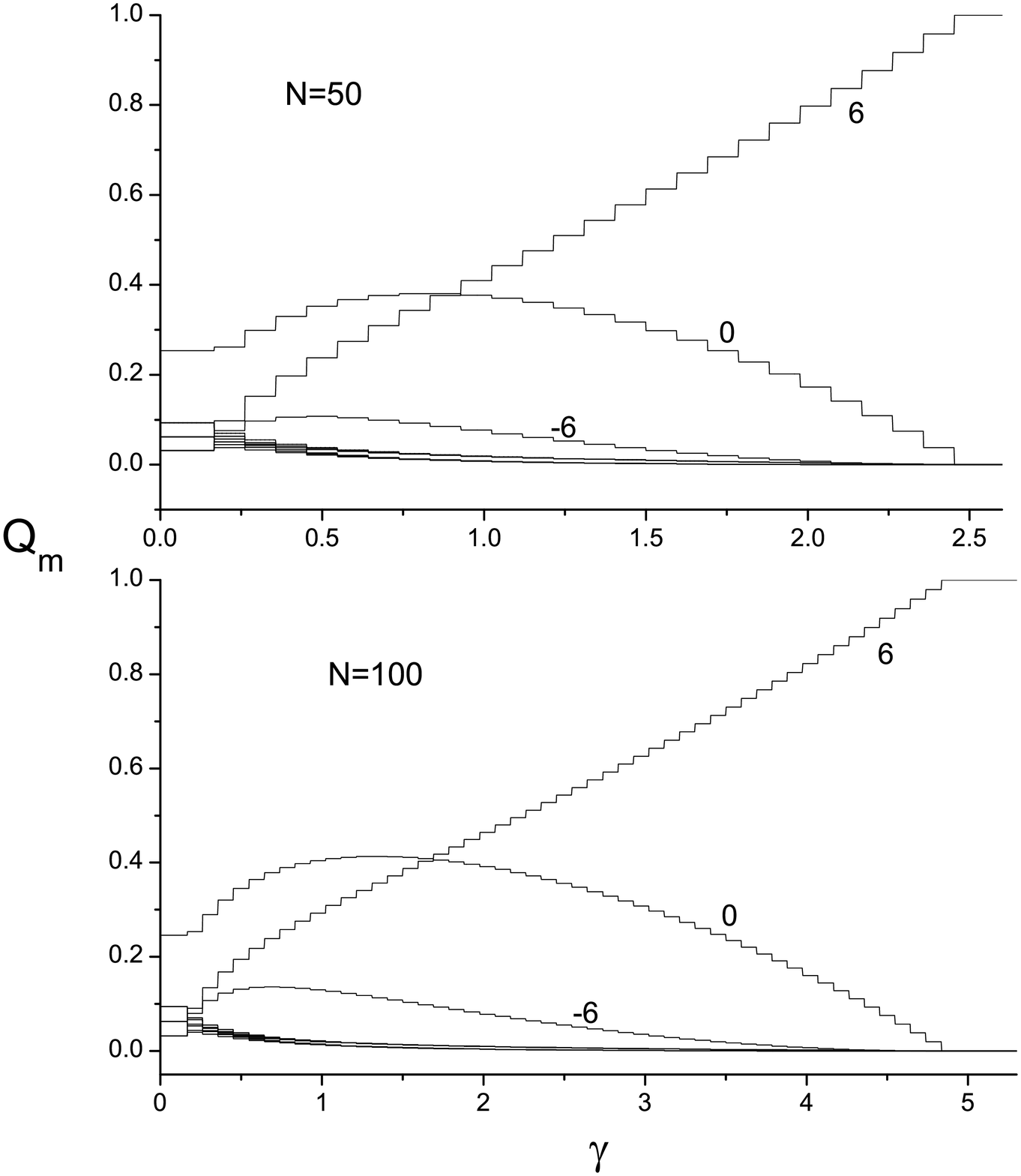}}
\caption{$Q_{m}$ (the probability of a pair of atoms with
magnetization $m$)
of the ground states of $H_{mod}^{\prime }$\ against $\protect\gamma $ with $%
N=50$ (a) and 100 (b). $m$ is marked by the associated curves (the lowest a
group of curves\ are not marked). $Q_{1}$ overlaps $Q_{2}$, $Q_{-1}$
overlaps $Q_{-2}$, $Q_{3}$ overlaps $Q_{4}$, and $Q_{-3}$ overlaps $Q_{-4}$.}
\label{fig:2}
\end{figure}

In addition to $P_{\mu }$, $Q_{m}$ is a measure of spin-correlation
(corresponding to the two-body correlation function in spatial space), and
has no counterpart in mean field theory. $Q_{m}$ of the ground states of $%
H_{mod}^{\prime }$ against $\gamma $\ are plotted in Fig.2. Among all the
thirteen $Q_{m}$\ ($m$ is from -6 to 6), $Q_{0}$ is much larger than the
others if $\gamma \leq 7/42$ (in this case all particles are in $s=0$
pairs). Thus, for every pair of particles, the probability of having their
spins anti-parallel is the largest. The next largest is the group $%
Q_{1}=Q_{2}=Q_{-1}=Q_{-2}$, while the smallest is the group $%
Q_{6}=Q_{5}=Q_{-6}=Q_{-5}$. However, when $\gamma $\ increases, $Q_{6}$ \
will increase nearly linearly. $Q_{0}$ and $Q_{-6}$ increase firstly with $%
\gamma $\ but eventually decrease with $\gamma $ as plotted. Other $Q_{m}$
are very small.\ When $\gamma \geq \gamma _{2}$, $Q_{6}=1$ implying a full
polarization as expected.

\section{Summary}

A model is proposed to study the possible pairing structures of N-boson
systems with nonzero spin. Analytical solutions have been obtained. The
following points are mentioned

When the pairing force is negative, the ground state has all particles
forming the $s=0$ pairs. The excited states are described by good quantum
numbers $S,\ M,$ and $I$. The latter is just the number of particles not in
the $s=0$ pairs. The state has a larger $I$ is higher, and the states have
the same $I$ are degenerate. When a magnetic field $B$ is applied, the
ground state will have a portion of particles fully polarized, while the
other particles remain in the $s=0$ pairs.

When the pairing force is positive, the ground state is highly degenerate
with a gap proportional to $N$. Therefore, when $N$\ is large, the
spin-degrees of freedom are difficult to be excited.

Two quantities $P_{\mu }$ and $Q_{m}$ have been defined and studied.
They are the spin-space analogues of the one- and two-body densities
of coordinate space. In particular, $Q_{m}$ can help us to
understand better the spin-correlation and has no counterpart in
mean field theory. Therefore, these quantities are in general useful
for understanding the spin-structures of various systems. For an
example, the underlying structure of the "polar phase" found in
\cite{DRB2006, SL2006, MH2007} is thereby clarified as a mixture of
fully polarized particles and $s=0$\ pairs.

In general, the results from this model system might be helpful for the
understanding of some realistic systems.

\begin{acknowledgements}
The support from the NSFC under the grant 10574163 and 10874249 is
appreciated.
\end{acknowledgements}


\begin{thebibliography}{99}
\bibitem{REF4} In the theory of condensates based on the mean field theory, $%
f(\mathbf{r)}$ depends on $\mu $, the Z-component of the spin of an atom.
When the dependence is omitted, it is called the single mode approximation
(SMA). This approximation would be valid if the particle density is dilute,
the temperature is very low, and the spin-dependent interaction is weak.
This is just the case of most condensates in general. Therefore, the SMA is
a good approximation and has been widely used in the literatures of
condensates. In this paper the SMA is used also.
\bibitem{REF5} Although the energy gap of the ground state does not depend
on $N$\ if $|G_{0}|$ is used as a unit, since $\int d\mathbf{r}|f(\mathbf{r}%
)|^{4}$ is roughly proportional to $N^{-3/5}$, $|G_{0}|$ decreases slowly
with the increase of $N$. Thus a larger $N$\ would actually lead to a
smaller gap when the pairing force is attractive.
\end{thebibliography}
\end{document}